\documentclass{winnower}
\newenvironment{proof}{\paragraph{Proof:}}{\hfill$\square$}
\usepackage{memhfixc}
\usepackage{graphicx}
\usepackage{lineno,hyperref}
\usepackage{amsmath}
\usepackage{latexsym}
\usepackage{amsfonts}
\usepackage{amssymb}
\usepackage{mathrsfs}
\usepackage[utf8]{inputenc}
\usepackage{bm}
\usepackage{graphicx}
\usepackage{hyperref}
\usepackage{subfigure}
\usepackage{pdfpages}
\newtheorem{theorem}{Theorem}[section]

\newtheorem{definition}{Definition}[section]
\newtheorem{lemma}[theorem]{Lemma}
\newtheorem{remark}{Remark}[section]
\newtheorem{corollary}[theorem]{Corollary}

\newcommand{\ba}{\begin{array}}
\newcommand{\ea}{\end{array}}

\newcommand{\bfl}{\begin{flushleft}}
\newcommand{\efl}{\end{flushleft}}
\newcommand{\bfr}{\begin{flushright}}
\newcommand{\efr}{\end{flushright}}
\newcommand{\bt}{\begin{theorem}}
\newcommand{\bd}{\begin{definition}}
\newcommand{\ed}{\end{definition}}
\newcommand{\et}{\end{theorem}}
\newcommand{\bl}{\begin{lemma}}
\newcommand{\el}{\end{lemma}}
\newcommand{\ee}{\end{exam}}
\newcommand{\bcor}{\begin{corollary}}
\newcommand{\ecor}{\end{corollary}}
\usepackage[utf8]{inputenc}
\hypersetup{colorlinks,linkcolor={red},citecolor={blue},urlcolor={red}} 
\usepackage{appendix}
\usepackage{fancyhdr}
 \usepackage{multirow}
\usepackage{tabularx}
\usepackage{makecell}
\usepackage{lscape}
\usepackage{float}
\usepackage{lipsum}
\usepackage{xcolor}
\usepackage{lipsum}
\makeatletter
\DeclareRobustCommand{\change}{%
  \@bsphack
  \leavevmode
  \color{red}%
  \@esphack
}
\DeclareRobustCommand{\stopchange}{%
  \@bsphack
  \normalcolor
  \@esphack
}
\makeatother

\begin{document}

\title{Two-sample nonparametric test for proportional reversed hazards}

\author{Ruhul Ali Khan}
\affil{Theoretical Statistics and Mathematics Unit, Indian Statistical Institute, Delhi Centre, New Delhi 110016, India}

\date{}

\maketitle

\let\thefootnote\relax\footnotetext{E-mail Addresses: Ruhul Ali Khan (ruhulali.khan@gmail.com)}

\begin{abstract}
Several works have been undertaken in the context of proportional reversed hazard rate (PRHR) since last few decades. But any specific statistical methodology for the PRHR hypothesis is absent in the literature. In this paper, a two-sample nonparametric test based on two independent samples is proposed for verifying the PRHR assumption. Based on a consistent U-statistic three statistical methodologies have been developed exploiting U-statistics theory, jackknife empirical likelihood and adjusted jackknife empirical likelihood method. A simulation study has been performed to assess the merit of the proposed test procedures. Finally, the test is applied to a data in the context of brain injury-related biomarkers and a data related to Ducheme muscular dystrophy.

{\bf Keywords:} 
Reversed hazard rate, Nonparametric test, Asymptotic normality, Empirical likelihood, Wilks’ theorem. \\

{\bf AMS Subject Classification:} Primary 62G10, Secondary 90B25.
\end{abstract}

%-------------------------------------------------%
\section{Introduction}\label{sec1}
%-------------------------------------------------%
The concept of reversed hazard rate plays a very prominent role in reliability and survival analysis. For a nonnegative random variable  $X$ associated with distribution function $F$, the reversed hazard
rate (RHR) is defined as
\begin{equation}
\label{rhrraw}
r(x)=\lim_{\Delta x \rightarrow 0}   \frac{P(x- \Delta x< X< x \vert X\leq x )}{\Delta x} .
\end{equation}
Thus, for a very small interval, the probability of failure in the interval given failure before the end of the interval is approximated by the product of the RHR and the length of the interval. If the probability density function $f$ exists then (\ref{rhrraw}) can be written as, $r(x)=f/F$. The concept of RHR was first introduced by \cite{von1936distribution} and was discussed briefly by \cite{barlow1963properties}. Later, \cite{keilson1982uniform} called it `dual failure function’ but the term `reversed hazard rate' first appeared explicitly in \cite{lagakos1988nonparametric}. The correspondence between the bivariate RHR and the bivariate df for arbitrary bivariate distributions with discontinuities can be found in \cite{dabrowska1988kaplan}. The usefulness of the property of RHR (sometimes called retro-hazard in survival analysis) is important in the estimation of the survival function under left censored data (see \cite{kalbfleisch1989inference}). The RHR has also been called retro-hazard in \cite{gross1992regression}. \cite{block1998reversed} discussed some properties of the RHR function while \cite{nanda2001hazard} derived some interesting results which compared order statistics in the RHR orders. \cite{chechile2011properties} pointed out that the hazard rate and RHR may look quite similar but they are very different by illustrating some
interesting examples. For application in medical studies and reliability analysis of RHR, one may refer to  \cite{kalbfleisch1991regression}, \cite{lawless2011statistical} and \cite{andersen2012statistical}.

Let $X$ and $Y$ be two nonnegative random variables having distribution functions $F_0$ and $F$ respectively. Then the proportional RHR model is defined by the following relationship between distribution functions
\begin{equation}
\label{resy}
F(x)=\left[ F_0(x) \right]^\theta, \quad \quad \theta>0,
\end{equation}
where $F_0$ is called the baseline distribution function. If the two probability density functions $f_0$ and $f$ exist, $r_0=f_0/F_0$ and $r=f/F$ are the reversed hazard rates corresponding to $F_0$ and $F$ respectively. \cite{marshall2007life} denoted $F$ as a resilience parameter family with underlying distribution $F_0$ and $\theta$ as the resilience parameter. Note that the relationship given by (\ref{resy}) corresponds to the proportional reversed hazard rate (PRHR) assumption since $r(t)=\theta r_0(t)$. \cite{gupta1998modeling} introduced proportional RHR model in contrast to the celebrated proportional hazard model (PHM) proposed by \cite{cox1972regression}. \cite{di2000some} and \cite{gupta2007proportional} investigated the proportional reversed hazards model and its applications. \cite{kundu2004characterizations} provided two simple characterizations of the PRHR class of distributions based on some conditional expectation and conditional variance. \cite{sengupta2010proportional} pointed out that PRHR is applicable where the PH model is inappropriate. \cite{wang2015interval} developed estimation methodology for the confidence intervals of the family of PRHR distributions based on lower record values. A proportional cause-specific reversed hazards model was introduced by \cite{sankaran2016proportional}.  
\cite{fallah2021statistical} proposed statistical  inference  for  component  lifetime  distribution  from  coherent  system  lifetimes  under  a PRHR  model. \cite{baratnia2021reversed} proposed a new statistical method for one-way classification which is based on the RHR function of the response variable and \cite{khan2021some} studied the relationship between RHR and mean inactive time function. Moreover, there exists many well known distributions which belong to the resilience  parameter  family. Some of these are the Burr X distribution (\cite{burr1942cumulative}), Topp-Leone distribution (\cite{topp1955family}),  generalized exponential distribution (\cite{gupta1999theory}), generalized Rayleigh distribution (\cite{kundu2005generalized}), generalized Gompertz distribution (\cite{el2013generalized}) and exponential-type distribution (\cite{lemonte2013new}). Ever since the inception of the RHR, due to its diverse applications, many researchers have contributed to it and a vast number of publications on this topic have appeared. Here, I have tried only to mention some significant works regarding RHR with a brief literature survey.

To the best of present author's knowledge, there does not exist any statistical methodology explicitly for the proportional reversed hazard hypothesis. In this paper, we construct a test procedure for testing

\begin{alignat}{2}
\label{testprob}
& H_0: \frac{r_F(t)}{r_{F_0}(t)} = \theta \text{ for all } t>0 &  & \nonumber \\
& vs. &   & \\
& H_1: \frac{r_F(t)}{r_{F_0}(t)} \text{ is an increasing function of } t>0 &  & \nonumber
\end{alignat}
where $\theta>0$ is an unknown constant. Note that if ${r_F(t)}/{r_{F_0}(t)}$ is a decreasing function of $t>0$. then we take take the alternative as $H_1^\prime: \frac{r_F(t)}{r_{F_0}(t)} \text{ is a decreasing function of } t>0$. %use ${r_{F_0}(t)}/{r_{F}(t)}$ instead of ${r_F(t)}/{r_{F_0}(t)}$.  

%\section{Testing problem}
%Let $X$ and $Y$ be two nonnegative random variables having distribution functions $F_0$ and $F$ respectively.  If the two probability density functions $f_0$ and $f$ exist, $r_0=f_0/F_0$ and $r=f/F$ are the reversed hazard rates corresponding to $F_0$ and $F$, respectively.  Note that $Y$ is smaller than $X$ in the convex transform order (denoted as $Y\prec_c X$ or $F\prec_c F_0$) when $F_0^{-1}F(x)$ is convex in $x$ on the support of $F$. Thus, following \cite{shaked2007stochastic}, it can be easily proved that $F\prec_c F_0$ if and only if $r/r_0$ is increasing on $[0, \infty)$ and $F$ is equal to $F_0$ in convex transform order if and only if $r(t)=\theta r_0 (t)$. This motivates to detect the proportional reversed hazard rate in a more meaningful manner. In the next section, we construct a test procedure for testing

%When $F_0\prec_c F$ the dual model is considered, we test  $H_0: \frac{r_F(t)}{r_{F_0}(t)} = \theta \text{ for all } t>0$ vs. $H_1: \frac{r_F(t)}{r_{F_0}(t)} \text{ is an increasing function of } t>0$. 

The rest of the paper is arranged as follows. Section \ref{sec2} deals with the formulation of the test procedure. There are three subsections of Section \ref{sec2}. In Subsection \ref{subsec2.1}, asymptotic normality of the test statistics has been established utilizing U-statistic theory. Motivated by the seminal work of \cite{jing2009jackknife}, we have also derived asymptotic results based on jackknife empirical likelihood method in Subsection \ref{subsec2.2}.  In Subsection \ref{subsec2.3}, we extend the result of Subsection \ref{subsec2.2} based on the adjusted jackknife empirical likelihood method. In Section \ref{sec3}, we assess the performance of the proposed three tests in various scenarios by means of a simulation study and the salient features are reported. Two real-life data sets have been analysed in Section \ref{sec4}. The first data in the context of a brain injury-related biomarker accepts the the null hypothesis of the proportionality in the reversed hazard rate. The second data related to Ducheme muscular dystrophy  rejects  the  null hypothesis in favour of the alternative. Finally, Section \ref{sec5} contains some concluding remarks and a discussion about possible avenues of future work.

\section{Formulation of the test procedure}
\label{sec2}
In this section, a non-parametric test will be developed for the testing problem given in (\ref{testprob}) based on the random samples $X_1, X_2, \dots , X_m$ of size $m$ from $F_0$ and $Y_1, Y_2, \dots , Y_n$ of size $n$ from $F$. Let $x_1, x_2 \in (0, \infty)$ and $x_1<x_2$. Then, under $H_1$, we have 
\begin{equation}
\label{diff}
F(x_2)f(x_1)F_0(x_1)f_0(x_2)\leq F(x_1)f(x_2)F_0(x_2)f_0(x_1)
\end{equation}
 with strict inequality for some $x_1$ and $x_2$. Now integrating the difference given in (\ref{diff}) over the range $x_1<x_2$ we obtain a measure of departure $\Delta(F, F_0)$ as 
 
 \begin{alignat}{2}
 \label{dep1}
& \Delta(F, F_0) & = & \iint\limits_{x_1<x_2}  \left[ F(x_1)f(x_2)F_0(x_2)f_0(x_1) - F(x_2)f(x_1)F_0(x_1)f_0(x_2)\right] dx_1 dx_2  \nonumber \\
& & = & \iint\limits_{x_1<x_2} F(x_1)f(x_2)F_0(x_2)f_0(x_1) dx_1 dx_2  - \iint\limits_{x_1<x_2} F(x_2)f(x_1)F_0(x_1)f_0(x_2) dx_1 dx_2 \nonumber \\
& & = & I_1 -I_2.
\end{alignat}
Now
 \begin{alignat}{2}
 \label{I1}
& I_1 & = & \iint\limits_{x_1<x_2} F(x_1)f(x_2)F_0(x_2)f_0(x_1) dx_1 dx_2   \nonumber \\
& & = & \int_0^\infty F(x_1) \left(\int_{x_1}^\infty F_0(x_2) dF(x_2) \right) d F_0 (x_1) \nonumber \\
& & = & \int_0^\infty F(x_1) P\left(X_1<Y_1; x_1<Y_1 \right) d F_0 (x_1) \nonumber \\
& & = & P(X_1<Y_1, Y_2<X_2<Y_1)
\end{alignat}
and 
 \begin{alignat}{2}
 \label{I2}
& I_2 & = & \iint\limits_{x_1<x_2} F(x_2)f(x_1)F_0(x_1)f_0(x_2) dx_1 dx_2  \nonumber \\
& & = & \int_0^\infty F_0(x_1) \left(\int_{x_1}^\infty F(x_2) dF_0(x_2) \right) d F (x_1) \nonumber \\
& & = & \int_0^\infty F_0(x_1) P\left(Y_1<X_1; x_1<X_1 \right) d F (x_1) \nonumber \\
& & = & P(Y_1<X_1, X_2<Y_2<X_1).
\end{alignat}
Now using (\ref{dep1}), (\ref{I1}) and (\ref{I2}), we get the following simplified measure of departure as
 \begin{equation}
 \label{departure}
 \Delta(F, F_0)= P(X_1<Y_1, Y_2<X_2<Y_1)- P(Y_1<X_1, X_2<Y_2<X_1),
 \end{equation}
where $X_1, X_2 \in F_0$, $Y_1, Y_2 \in F$ are four independent random variables. It is clear that  $ \Delta(F, F_0)=0$ under the null hypothesis and $ \Delta(F, F_0)>0 (<0)$ under the alternative $H_1$ ($H_1^\prime$). Define a kernel
 \begin{equation}
 \label{kerunsym}
 f(X_1, X_2, Y_1, Y_2)= I(X_1<Y_1, Y_2<X_2<Y_1)- I(Y_1<X_1, X_2<Y_2<X_1),
 \end{equation}
 where $I$ is an indicator function. Then, $E\left( f(X_1, X_2, Y_1, Y_2) \right)= \Delta(F, F_0)$. Let $\phi(X_1, X_2, Y_1, Y_2)$ be the symmetric version of the kernel $f(X_1, X_2, Y_1, Y_2)$, i,e.,
 
 \begin{alignat}{2}
 \label{symker}
& \phi(X_1, X_2, Y_1, Y_2) & = & \frac{1}{4}  \Big[ I(X_1<Y_1, Y_2<X_2<Y_1)- I(Y_1<X_1, X_2<Y_2<X_1)   \nonumber \\
& & & + I(X_2<Y_1, Y_2<X_1<Y_1)- I(Y_1<X_2, X_1<Y_2<X_2) \nonumber \\
& & & + I(X_1<Y_2, Y_1<X_2<Y_2)- I(Y_2<X_1, X_2<Y_1<X_1) \nonumber \\
& & & +  I(X_2<Y_2, Y_1<X_1<Y_2)- I(Y_2<X_2, X_1<Y_1<X_2)\Big].
\end{alignat} 
 Then, an unbiased and consistent estimator of $\Delta(F, F_0)$ is given by the U-statistic
 \begin{equation}
 \label{ustattis}
 U= {m \choose 2}^{-1} {n \choose 2}^{-1} \mathop{{\sum \sum}}_{1\leq i< j \leq m} \,\,\,\, \mathop{{\sum \sum}}_{1\leq k< l \leq n} \phi(X_i, X_j, Y_k, Y_l).
 \end{equation}
%Note that,
%\begin{equation}
%\phi(x_1, x_2, y_1, y_2)=
%\begin{cases}
% \,\,\,\, 2 & \text{ for Case 1}\\
% \,\,\,\, 1 & \text{ for Case 2}\\
% \,\,\,\, 0 & \text{ for Case 3 or Case 4}\\
% -1 &  \text{ for Case 5}\\
% -2 & \text{ for Case 6}
%\end{cases}
%\end{equation}
%where 
%\begin{alignat}{2}
%\label{testprob}
%& \text{Case 1: } & = & a  \nonumber \\
%& \text{Case 2: } & = & b \nonumber \\
%& \text{Case 3: } & = & c \nonumber\\
%& \text{Case 1: } & = & a  \nonumber \\
%& \text{Case 2: } & = & b \nonumber \\
%& \text{Case 3: } & = & c \nonumber\\
%& \text{Case 3: } & = & c \nonumber
%\end{alignat}

\subsection{Asymptotic results based on two sample U-statistic theory}
\label{subsec2.1}
The test procedure is to reject the null hypothesis $H_0$ in favour of the alternative hypothesis for large values of $U$. As the test statistic given in (\ref{ustattis}) is a U-statistic it is consistent and has asymptotically normal distribution. The following theorem follows from \cite{lehmann1951consistency}.

\begin{theorem}
Under $H_1$, as $n\rightarrow \infty$, $U$ converges in probability to $\Delta(F,F_0)$.
\end{theorem}

Now the asymptotic distribution of $U$ will be derived exploiting $U$-statistic theory. Let $m, n \rightarrow \infty$ in such a way that $m/(m + n) \rightarrow p$, $0 < p < 1$. Define
\begin{equation}
\label{sig10}
\sigma_{10}^2= Cov\left[\phi(X_1, X_2, Y_1, Y_2), \phi(X_1, X_3, Y_3, Y_4)\right]
\end{equation}
and
\begin{equation}
\label{sig01}
\sigma_{01}^2= Cov\left[\phi(X_1, X_2, Y_1, Y_2), \phi(X_3, X_4, Y_1, Y_3)\right].
\end{equation}

Now following Theorem 3.4.8 of \cite{randles1979introduction}[page 91] or \cite{lee1990u}, the asymptotic distribution of $U$ has been derived in the next theorem.
\begin{theorem}
\label{uallthm}
If $E\left[\phi^2(X_1, X_2, Y_1, Y_2)\right]<\infty $, then $\sqrt{m+n}\left[ U-\Delta(F, F_0)\right]$ converges in distribution to Gaussian random variable with mean zero and variance $\sigma^2$ as $\min(m,n)\rightarrow\infty$ where
\begin{equation}
\sigma^2=4\left(\frac{\sigma_{10}^2}{p} + \frac{\sigma_{01}^2}{1-p}\right).
\end{equation}
\end{theorem}
%A lengthy and involved algebraic manipulation yields the following
 Now we will proceed to derive the limiting distribution under the null hypothesis of the proportionality in the reversed hazard rate. 

\begin{theorem}
Assume that the conditions of Theorem \ref{uallthm} hold. Then, under $H_0$, 
$$\frac{U}{2\sqrt{\frac{\sigma_{10}^2(\theta)}{m}+ \frac{\sigma_{01}^2(\theta)}{n}}}\xrightarrow[]{d} N(0, 1) \,\, \text{ as } \min(m,n)\rightarrow\infty,$$
where
\begin{equation}
\label{sig10nul}
\sigma_{10}^2(\theta)= \frac{1}{4(2\theta+1)}\left[ 1 - \frac{1}{(2\theta+1)(\theta +1)^2} -\frac{8\theta}{3\theta + 2} + \frac{16 \theta^2}{(4 \theta +3)(2\theta +1)} \right]
\end{equation}
and
\begin{equation}
\label{sig01nul}
\sigma_{01}^2(\theta)= \frac{\theta}{4(2+\theta)}\left[1-\frac{8}{3+2\theta}+\frac{16}{(4+3\theta)(2+\theta)}-\frac{\theta^3}{(2+\theta)(\theta+1)^2} \right].
\end{equation}
\end{theorem}
\begin{proof}
Note that $\Delta(F, F_0)=0$ under $H_0$. Thus, following Theorem \ref{uallthm}, it is enough to show that the null asymptotic variance is $4\left(\frac{\sigma_{10}^2(\theta)}{p} + \frac{\sigma_{01}^2(\theta)}{1-p}\right)$ where $\sigma_{10}^2(\theta)=E_{X_1}\left[E\left\lbrace \phi(X_1, X_2, Y_1, Y_2)\vert X_1 \right\rbrace \right]^2$ and $\sigma_{01}^2(\theta)=E_{Y_1}\left[E\left\lbrace \phi(X_1, X_2, Y_1, Y_2)\vert Y_1 \right\rbrace \right]^2$. A lengthy and involved algebraic manipulation yields

\begin{equation}
\label{seventin}
E\left\lbrace \phi(X_1, X_2, Y_1, Y_2)\vert Y_1 \right\rbrace = \frac{1}{2}\left[\frac{\theta^2}{(2+\theta)(1+\theta)}-F^{1/\theta}(Y_1) +\frac{4}{2+\theta} F^{1+2/\theta}(Y_1) \right].   
\end{equation}
Thus the expression of $\sigma_{01}^2(\theta)$ given in (\ref{sig01nul}) has obtained from (\ref{seventin}) and  the fact that $\sigma_{01}^2(\theta)=E_{Y_1}\left[E\left\lbrace \phi(X_1, X_2, Y_1, Y_2)\vert Y_1 \right\rbrace \right]^2$. The expression of $\sigma_{10}^2(\theta)$ has been obtained from the relation that $\sigma_{10}^2(\theta)=\sigma_{01}^2(1/\theta)$. 
\end{proof}
\\

If $\theta$ is not specified in the null hypothesis, one can substitute a consistent estimator of $\theta$ in the expressions given in (\ref{sig10nul}) and (\ref{sig01nul}) in order to apply the test. Note that $\tau=P(X<Y)=\frac{\theta}{1+\theta}$ and a consistent estimator of $\tau$ can be obtained using Mann-Whitney statistics,
$$\hat{\tau}^{MW}=\frac{1}{mn}\sum_{i=1}^m \sum_{j=1}^n I(X_i< Y_i)$$
where $I(x < y)=1$ if $x < y$, $0$ otherwise. Thus a consistent estimator of $\theta$ is given by $\hat{\theta}^{MW}=\frac{\hat{\tau}^{MW}}{1-\hat{\tau}^{MW}}$ when $\hat{\tau}^{MW}\neq 1$. 
\begin{remark}
An application of Jensen inequality and the fact that $g(x)=\frac{x}{1-x}$ is a convex function on $(0,1)$ yields $E(\hat{\theta}^{MW})\geq \theta$. Thus $\hat{\theta}^{MW}$ is a biased estimator for $\theta$ but $\hat{\tau}^{MW}$ is an unbiased estimator for $\tau$.  
\end{remark}
The symbol $U_{MW}$ will be used to represent the U-statistic which is normalised as the following manner.
Denote 
\begin{equation}
U_{MW}=\frac{U}{2\sqrt{\frac{\sigma_{10}^2(\hat{\theta}^{MW})}{m}+ \frac{\sigma_{01}^2(\hat{\theta}^{MW})}{n}}} \end{equation}
Thus, one can reject the null hypothesis
of proportionality of reversed hazard rate in favour of the alternative $H_1$ ($H_1^\prime$) if $U_{MW}\geq Z_{\alpha} (\leq -Z_{\alpha})$ for large values of $(m+n)$, where $Z_{\alpha}$ is the upper $\alpha$-th quantile of the standard normal distribution.

\subsection{Asymptotic results based on jackknife empirical likelihood}
\label{subsec2.2}

The empirical likelihood (EL) is widely applied in various statistical situations as a nonparametric technique since it combines the robustness of nonparametric methods while enjoying the flexibility and effectiveness of likelihood approach as parametric methods. EL was first used by \cite{thomas1975confidence} to construct better confidence intervals involving the Kaplan-Meier estimators in survival analysis. Later, some seminal works of \citeauthor{owen1988empirical} (\citeyear{owen1988empirical}, \citeyear{owen1990empirical}, \citeyear{owen1991empirical}, \citeyear{owen2001empirical}) established various important and interesting properties of this method. One of the most appealing property is that the asymptotic distribution of the empirical likelihood ratio test statistic follows a chi-squared distribution, which is the same as the one under parametric settings. However, when this method gets involved with some nonlinear statistics such as U-statistics, its computational advantage may vanish due to the increasing difficulties in solving a number of nonlinear equations simultaneously by Lagrange multiplier method. \cite{jing2009jackknife} gave an example to illustrate this situation. \cite{shi1984approximate} established that the jackknife pseudo-values are asymptotically independent under mild conditions which was posed as a conjecture by \cite{tukey1958bias}. Exploiting the result of \cite{shi1984approximate}, \cite{jing2009jackknife} proposed a jackknife empirical likelihood (JEL) method which combines
jackknife and the Owen’s empirical likelihood approaches since the statistic of interest can be written as a sample mean based on jackknife pseudo-values (\cite{quenouille1956notes}). The proposed method of \cite{jing2009jackknife} is simple and the Wilks’ theorem was also established in this context.  In this paper, the proposed method of \cite{jing2009jackknife} will be adapted for the purposes of comparison. 

The test statistic given in (\ref{ustattis}) can be
expressed as

\begin{alignat}{2}
\label{jingstat}
& U & = & {m \choose 2}^{-1} {n \choose 2}^{-1} \mathop{{\sum \sum}}_{1\leq i< j \leq m} \,\,\,\, \mathop{{\sum \sum}}_{1\leq k< l \leq n} \phi(X_i, X_j, Y_k, Y_l)\nonumber \\
& & := & T(X_1, X_2, \dots , X_m, Y_1, Y_2, \dots , Y_n).
\end{alignat}
where $\Delta(F, F_0) = E\left(\phi(X_1, X_2, Y_1, Y_2) \right)$ is a parameter of interest. 
Denote $\bm{Z}$ as the combined samples of $\bm{X}$ and $\bm{Y}$ where  
\begin{equation}
Z_i=
\begin{cases}
X_i & \text{ for } i=1,2,\dots , m\\
Y_{i-m}  & \text{ for } i=(m+1), (m+2),\dots , (m+n).
\end{cases}
\end{equation}
Then, (\ref{jingstat}) can be written as
\begin{equation}
U= T(X_1, X_2, \dots , X_m, Y_1, Y_2, \dots , Y_n)=T(Z_1, \dots , Z_{m+n}).   
\end{equation}
Now the jackknife pseudo-values are given by
\begin{equation}
 \hat{V_i}= (m+n)T-(m+n-1)T^{(-i)},   
\end{equation}
where $T^{(-i)}$ is the computed value of $T$ based on the observations $Z_1, \dots, Z_{i-1}, Z_{i+1}, \dots, Z_{m+n}$. Thus it can be easily shown that $T=\frac{1}{m+n}\sum_i^{m+n} \hat{V_i}$. In general, $\hat{V_i}$’s are dependent r.v.’s, but asymptotically independent under mild conditions (see \cite{shi1984approximate}). Now, the JEL method exploiting Owen’s empirical likelihood is as follows.

Let $p = (p_1, \dots , p_{m+n})$ be a probability vector, assigning probability $p_i$ to $\hat{V_i}$. The JEL, evaluated at $\Delta$, becomes
\begin{equation}
 L(\Delta)=\max \left\lbrace \prod_{i=1}^{m+n} p_i : \sum_{i=1}^{m+n} p_i=1, \sum_{i=1}^{m+n} p_i \left(\hat{V_i} -E\hat{V_i} \right)=0 \right\rbrace ,  
\end{equation}
where
\begin{equation}
E\hat{V_i}=
\begin{cases}
\frac{\Delta(m+n)(2n-m-2)}{(m+n-4)m}  & \text{ for } i=1,2,\dots , m\\
\frac{\Delta(m+n)(2m-n-2)}{(m+n-4)n}  & \text{ for } i=(m+1), (m+2),\dots , (m+n).
\end{cases}
\end{equation}
Note that $\prod_{i=1}^{m+n} p_i$ subject to $\sum_{i=1}^{m+n} p_i=1$, attains its maximum $(m+n)^{-(m+n)}$ at $p_i = 1/(m+n)$. So we define the jackknife empirical likelihood ratio at $\Delta$ by
\begin{alignat}{2}
\label{jelratio}
& R(\Delta) & = & \frac{L(\Delta)}{(m+n)^{-(m+n)}} \nonumber\\
& & = & \max \left\lbrace \prod_{i=1}^{m+n} ((m+n)p_i) : \sum_{i=1}^{m+n} p_i=1, \sum_{i=1}^{m+n} p_i \left(\hat{V_i} -E\hat{V_i} \right)=0 \right\rbrace .  
\end{alignat}
Using Lagrange multipliers, when $\displaystyle{\min_{1\leq i\leq m+n}}\left(\hat{V_i} -E\hat{V_i} \right) < 0 < \displaystyle{\max_{1\leq i\leq m+n}}\left(\hat{V_i} -E\hat{V_i} \right)$, we have

$$p_i=\frac{1}{m+n}\frac{1}{1+\lambda(\hat{V_i} -E\hat{V_i})}$$ 
where $\lambda=\lambda(\Delta)$ satisfies
\begin{equation}
\label{lambdasol}
\frac{1}{m+n}\sum_{i=1}^{m+n} \frac{\hat{V_i} -E\hat{V_i}}{1+\lambda(\hat{V_i} -E\hat{V_i})}=0.    
\end{equation}

After plugging the $p_i$’s back into (\ref{jelratio}) and taking the logarithm of $R(\Delta)$, the nonparametric jackknife empirical log-likelihood ratio is given by
\begin{equation}
\log{R(\Delta)}=-\sum_{i=1}^{m+n} \log{\left[1+\lambda(\hat{V_i}-E\hat{V_i})\right]}.
\end{equation}

Following Theorem 2 of \cite{jing2009jackknife}, we have the following theorem which is an analogue of Wilks' theorem.

\pagebreak

\begin{theorem} 
\label{jelthm}
Assume that
\begin{enumerate}
    \item $E\left[\phi^2(X_1, X_2, Y_1, Y_2)\right]<\infty $,
    \item $\sigma_{10}$ and $\sigma_{01}$ defined in (\ref{sig10}) and (\ref{sig01}), respectively, are positive
    \item and $0< \lim\inf \frac{m}{n}\leq \lim\sup \frac{m}{n}<\infty$.
\end{enumerate}
Then, under $H_0$, $-2 \log(R(\Delta))$ converges in distribution to $\chi^2$ random variable with one degree of freedom.
\end{theorem}

Thus in view of Theorem \ref{jelthm}, we reject the null hypothesis $H_0$ against the alternative $H_1$ ($H_1^\prime$) at a level of significance $\alpha$, if
$$U_{JEL}:=-2 \log(R(0))>\chi^2_{1, \alpha} (< -\chi^2_{1, \alpha}),$$
where $\chi^2_{1, \alpha}$ is the $1-\alpha$ quantile of the $\chi^2$ distribution with one degree of freedom.

\subsection{Asymptotic results based on adjusted jackknife empirical likelihood}
\label{subsec2.3}
When the sample size is small, the convex hull of $\hat{V_i}-E\hat{V_i}$ may not contain $0$. As a result, (\ref{lambdasol}) might not always have a solution. To make sure the constrained maximization always has a solution, using the concept of adjusted empirical likelihood proposed by \cite{chen2008adjusted}, adjusted jackknife empirical likelihood (AJEL) was developed by \cite{chen2016adjusted}. We adapt their approach to the JEL by adding one more jackknife pseudo-value which is defined as
$$\hat{V}_{m+n+1}=-\frac{a_{n+m}}{m+n}\sum_{i=1}^{m+n} \hat{V}_i.$$
\cite{chen2008adjusted} suggested at $a_{n+m} = \max(1, \log{(m+n)}/2)$ and applying empirical likelihood method to these $m+n+1$ jackknife pseudo-values values,
the adjusted jackknife empirical likelihood ratio at $\Delta$ is defined as follows (cf. \cite{lin2017jackknife}),

\begin{equation}
\label{ajelratio}
R^*(\Delta) =  \max \left\lbrace \prod_{i=1}^{m+n+1} ((m+n+1)p_i) : \sum_{i=1}^{m+n+1} p_i=1, \sum_{i=1}^{m+n+1} p_i \left(\hat{V_i} -E\hat{V_i} \right)=0 \right\rbrace .  
\end{equation}
The adjusted jackknife empirical log-likelihood ratio is
\begin{equation}
\log{R^*(\Delta)}=-\sum_{i=1}^{m+n+1} \log{\left[1+\lambda(\hat{V_i}-E\hat{V_i})\right]}.
\end{equation}
where the Lagrange multiplier $\lambda=\lambda(\Delta)$ satisfies
\begin{equation}
\label{lambdasolad}
\frac{1}{m+n+1}\sum_{i=1}^{m+n+1} \frac{\hat{V_i} -E\hat{V_i}}{1+\lambda(\hat{V_i} -E\hat{V_i})}=0.    
\end{equation}
With the same conditions given by \cite{jing2009jackknife}, Wilks theorem of the AJEL has been established by \cite{chen2016adjusted}. Thus, as a special case, the following
theorem holds for the above AJEL ratio. For the proof, we refer the reader to
\cite{chen2016adjusted}.

\begin{theorem} 
\label{ajelthm}
Under the assumptions of Theorem \ref{jelthm}, $-2 \log(R^*(\Delta))$ converges in distribution to $\chi^2$ random variable with one degree of freedom.
\end{theorem}

Thus in view of Theorem \ref{ajelthm}, we reject the null hypothesis $H_0$ against the alternative $H_1$ ($H_1^\prime$) at a level of significance $\alpha$, if
$$U_{AJEL}:=-2 \log(R^*(0))>\chi^2_{1, \alpha} (< -\chi^2_{1, \alpha}),$$
where $\chi^2_{1, \alpha}$ is the $(1-\alpha)$ quantile of the $\chi^2$ distribution with one degree of freedom.

\begin{table}
\setlength\tabcolsep{4pt}
\centering
\caption{Empirical Type I error of the proposed tests based on $U_{MW}$, $U_{JEL}$ and $U_{AJEL}$}
\label{sizeprhr}
\begin{tabular}{lcccclccclccc} 
\hline
\multicolumn{1}{c}{\multirow{2}{*}{\begin{tabular}[c]{@{}c@{}}\\ $\theta$\end{tabular}}} & \multirow{2}{*}{$(m, n)$} & \multicolumn{3}{c}{$U_\text{MW}$}                                              & \multicolumn{1}{c}{} & \multicolumn{3}{c}{\begin{tabular}[c]{@{}c@{}}$U_\text{JEL}$ \\\end{tabular}}        & \multicolumn{1}{c}{} & \multicolumn{3}{c}{$U_\text{AJEL}$}                                                   \\ 
\cline{3-5}\cline{7-9}\cline{11-13}
\multicolumn{1}{c}{}                                                                           &                           & $\alpha$=0.01           & $\alpha$=0.05                 & $\alpha$=0.10                 &                      & $\alpha$=0.01                 & $\alpha$=0.05                 & $\alpha$=0.10                 &                      & $\alpha$=0.01                 & $\alpha$=0.05                 & $\alpha$=0.10                  \\ 
\hline
\multicolumn{1}{c}{2}                                                                          & (10, 10)                  & 0.0179               & 0.0951                     & 0.1630                     & \multicolumn{1}{c}{} & 0.0425                     & 0.1094                     & 0.1624                     &                      & 0.0204                     & 0.0837                     & 0.1333                      \\
                                                                                               & (20, 20)                  & 0.0137               & \multicolumn{1}{l}{0.0668} & \multicolumn{1}{l}{0.1255} &                      & 0.0301                     & \multicolumn{1}{l}{0.0641} & \multicolumn{1}{l}{0.1059} &                      & \multicolumn{1}{l}{0.0254} & \multicolumn{1}{l}{0.0565} & \multicolumn{1}{l}{0.0922}  \\
                                                                                               & (25, 20)                  & 0.0118               & 0.0644                     & 0.1222                     &                      & \multicolumn{1}{l}{0.0245} & \multicolumn{1}{l}{0.0633} & \multicolumn{1}{l}{0.1092} &                      & \multicolumn{1}{l}{0.0204} & \multicolumn{1}{l}{0.0554} & \multicolumn{1}{l}{0.0946}  \\
                                                                                               & \multicolumn{1}{l}{}      & \multicolumn{1}{l}{} & \multicolumn{1}{l}{}       & \multicolumn{1}{l}{}       &                      & \multicolumn{1}{l}{}       & \multicolumn{1}{l}{}       & \multicolumn{1}{l}{}       &                      & \multicolumn{1}{l}{}       & \multicolumn{1}{l}{}       & \multicolumn{1}{l}{}        \\
\multicolumn{1}{c}{4}                                                                          & (10, 10)                  & 0.0040               & \multicolumn{1}{l}{0.0779} & \multicolumn{1}{l}{0.1697} &                      & 0.0207                     & 0.1150                     & 0.1908                     &                      & 0.0048                     & 0.0680                     & 0.1436                      \\
                                                                                               & (20, 20)                  &   0.0118                   &  0.0726                          & 0.1387                            &                      &  0.0631                          &     0.1173                       &   0.1626                         &                      &    0.0495                        &    0.1058                        &   0.1463                          \\
                                                                                               & (25, 20)                  &   0.0070                   &   0.0624                         &   0.1219                         &                      &  0.0511                          &   0.0975                         & 0.1388                           &                      &    0.0417                        &   0.0887                         &  0.1265                           \\
                                                                                               & \multicolumn{1}{l}{}      & \multicolumn{1}{l}{} & \multicolumn{1}{l}{}       & \multicolumn{1}{l}{}       &                      & \multicolumn{1}{l}{}       & \multicolumn{1}{l}{}       & \multicolumn{1}{l}{}       &                      & \multicolumn{1}{l}{}       & \multicolumn{1}{l}{}       & \multicolumn{1}{l}{}        \\
\multicolumn{1}{c}{6}                                                                          & (10, 10)                  & 0.0023               & 0.0611                     & \multicolumn{1}{l}{0.1603} &                      & 0.0110                     & 0.0952                     & 0.1990                     &                      & 0.0024                     & 0.0465                     & 0.1328                      \\
                                                                                               & (20, 20)                  &   0.0056                      &   0.0694                         &  0.1475                          &                    &   0.0766                         &   0.1572                         &  0.2088                           &                      &  0.0551                          &   0.1399                         &  0.1935                           \\
                                                                                               & (25, 20)                  &  0.0036                    & 0.0546                            &  0.1241                          &                      & 0.0703                            &   0.1320                         &   0.1813                         &                     &     0.0552                        &  0.1190                          &  0.1684                           \\
\hline
\end{tabular}
\end{table}

\section{Simulation study}
\label{sec3}
In this section, Monte Carlo simulation studies have been carried out to assess the finite sample performance of the tests based on $U_{MW}$, $U_{JEL}$ and $U_{AJEL}$. The simulations are performed in R statistical software (\cite{asdde}) on PC platform. The empirical type I error and power of the tests based on $U_{MW}$, $U_{JEL}$ and $U_{AJEL}$ are calculated using the ``emplik" package of \cite{empzhou}.

In order to obtain empirical Type I error, we consider generalized exponential distribution (GED) which was introduced by \cite{gupta1999theory}. The cumulative distribution function (cdf) of GED is given by 
\begin{equation}
F(t)=\left(1-e^{-\lambda t}\right)^\theta, \quad t>0,
\end{equation}
where $\lambda>0$ and $\theta>0$ are the scale and resilience parameters  respectively and denoted by GED$(\lambda,\theta)$.  Note that GED belongs to the PRHR family with underlying distribution $F_0(t)=1-e^{-\lambda t}$, i.e., the exponential distribution. We generate two sets of independent random observations from exponential distribution with mean $1/\lambda$ of size $m$ and GED$(\lambda,\theta)$ of size $n$ respectively. The values of $\theta= 2, 4, 6$ and $\lambda=1$ were considered for the simulation study. The empirical type I error of the tests based on $U_{MW}$, $U_{JEL}$ and $U_{AJEL}$ are calculated based on 10,000 replications with sample sizes $(m,n)=(10, 10), (20, 20) \text{ and } (25, 20)$. We compute the proportion of times the null hypothesis is rejected for 1\%, 5\% and 10\% levels of significance. The results are reported in Table \ref{sizeprhr} which indicates that the empirical type I error approaches to the specified level of significance when the sample size increases or the value of $\theta$ increases.

\begin{table}
\setlength\tabcolsep{4pt}
\centering
\caption{Empirical power for Scenario I}
\label{powfrech}
\begin{tabular}{lcccclccclccc} 
\hline
\multicolumn{1}{c}{\multirow{2}{*}{\begin{tabular}[c]{@{}c@{}}\\ $a_2$\end{tabular}}} & \multirow{2}{*}{$(m, n)$} & \multicolumn{3}{c}{$U_\text{MW}$}                                  & \multicolumn{1}{c}{} & \multicolumn{3}{c}{\begin{tabular}[c]{@{}c@{}}$U_\text{JEL}$ \\\end{tabular}} & \multicolumn{1}{c}{} & \multicolumn{3}{c}{$U_\text{AJEL}$}                                 \\ 
\cline{3-5}\cline{7-9}\cline{11-13}
\multicolumn{1}{c}{}                                                                        &                           & $\alpha$=0.01           & $\alpha$=0.05           & $\alpha$=0.05           &                      & $\alpha$=0.01           & $\alpha$=0.05           & $\alpha$=0.10                      &                      & $\alpha$=0.01          & $\alpha$=0.05           & $\alpha$=0.10            \\ 
\hline
\multicolumn{1}{c}{3}                                                                       & (10, 10)                  & 0.5333               & 0.7898               & 0.8730               & \multicolumn{1}{c}{} & 0.4480               & 0.6193               & 0.7027                          &                      & 0.3630               & 0.5672               & 0.6628                \\
                                                                                            & (20, 20)                  & 0.8658               & 0.9658               & 0.9846               &                      & 0.7519               & 0.8924               & 0.9420                          &                      & 0.7182               & 0.8743               & 0.9313                \\
                                                                                            & (25, 20)                  & 0.8976               & 0.9736               & 0.9901               &                      & 0.8044               & 0.9255               & 0.9594                          &                      & 0.7723               & 0.9117               & 0.9517                \\
                                                                                            & \multicolumn{1}{l}{}      & \multicolumn{1}{l}{} & \multicolumn{1}{l}{} & \multicolumn{1}{l}{} &                      & \multicolumn{1}{l}{} & \multicolumn{1}{l}{} & \multicolumn{1}{l}{}            &                      & \multicolumn{1}{l}{} & \multicolumn{1}{l}{} & \multicolumn{1}{l}{}  \\
\multicolumn{1}{c}{5}                                                                       & (10, 10)                  & 0.8210               & 0.9499               & 0.9750               &                      & 0.7286               & 0.8594               & 0.9108                          &                      & 0.6254               & 0.8206               & 0.8869                \\
                                                                                            & (20, 20)                  & 0.9902               & 0.9992               & 0.9997               &                      & 0.9728               & 0.9933               & 0.9975                          &                      & 0.9651               & 0.9916               & 0.9969                \\
                                                                                            & (25, 20)                  & 0.9950               & 0.9993               & 0.9998               &                      & 0.9811               & 0.9960               & 0.9986                          &                      & 0.9752               & 0.9953               & 0.9982                \\
                                                                                            & \multicolumn{1}{l}{}      & \multicolumn{1}{l}{} & \multicolumn{1}{l}{} & \multicolumn{1}{l}{} &                      & \multicolumn{1}{l}{} & \multicolumn{1}{l}{} & \multicolumn{1}{l}{}            &                      & \multicolumn{1}{l}{} & \multicolumn{1}{l}{} & \multicolumn{1}{l}{}  \\
\multicolumn{1}{c}{7}                                                                       & (10, 10)                  & 0.9144               & 0.9819               & 0.9915               &                      & 0.8335               & 0.9247               & 0.9562                          &                      & 0.7209               & 0.8974               & 0.9395                \\
                                                                                            & (20, 20)                  & 0.9989               & 1.0000               & 1.0000               &                      & 0.9959               & 0.9996               & 1.0000                          &                      & 0.9929               & 0.9992               & 0.9999                \\
                                                                                            & (25, 20)                  & 1.0000               & 1.0000               & 1.0000               &                      & 0.9996               & 1.0000               & 1.0000                          &                      & 0.9991               & 1.0000               & 1.0000                \\
\hline
\end{tabular}
\end{table}

We evaluate the empirical power of the proposed test  based on $U_{MW}$, $U_{JEL}$ and $U_{AJEL}$ by considering the following two scenarios:

{\bf Scenario I:} In this scenario, we have considered the Fr\'echet distribution, also known as inverse Weibull distribution. The cumulative distribution function of the Fr\'echet distribution is given by
\begin{equation}
\label{frech}
F(x)=e^{-x^{-\alpha}}, \quad x>0,
\end{equation}
where $\alpha>0$ is a shape parameter. Let $X$ and $Y$ follow the Fr\'echet distributions with shape parameter $\alpha_2$ and $\alpha_1$ respectively. Note that $\frac{r_Y(x)}{r_{X}(x)} \text{ is an increasing function of } x>0$ if and only if $\alpha_2>\alpha_1$. The values of $\alpha_2= 3, 5, 7$ and $\alpha_1=1$ were considered and we generate two sets of independent random observations of size $m$ and $n$ respectively from the Fr\'echet distributions
with respective shape parameter $\alpha_2$ and $\alpha_1$. The empirical power of the tests based on $U_{MW}$, $U_{JEL}$ and $U_{AJEL}$ are calculated based on 10,000 replications with sample sizes $(m,n)=(10, 10), (20, 20) \text{ and } (25, 20)$. The computed empirical power is presented in Table \ref{powfrech} for 1\%, 5\% and 10\% levels of significance. Table \ref{powfrech} suggests that all the proposed tests have good power for the parameters when $m\geq 20$ and $n\geq 20$ and the test $U_{MW}$ performs slightly better than $U_{JEL}$ for small values of $m$ and $n$. Moreover, $U_{JEL}$ has also marginally better power for the chosen parameters when compared with the test $U_{AJEL}$ for small values of $m$ and $n$. Finally, as expected the empirical power increases when the difference between $a_1$ and  $a_2$ increases.

\begin{table}[ht]
\setlength\tabcolsep{4pt}
\centering
\caption{Empirical power for Scenario II}
\label{powgum}
\begin{tabular}{lclccclccclccc} 
\hline
\multicolumn{1}{c}{\multirow{2}{*}{\begin{tabular}[c]{@{}c@{}}\\$\gamma$\end{tabular}}} & \multirow{2}{*}{$(m, n)$} &                              & \multicolumn{3}{c}{$U_\text{MW}$}                                  & \multicolumn{1}{c}{} & \multicolumn{3}{c}{\begin{tabular}[c]{@{}c@{}}$U_\text{JEL}$ \\\end{tabular}} & \multicolumn{1}{c}{} & \multicolumn{3}{c}{$U_\text{AJEL}$}                                 \\ 
\cline{4-6}\cline{8-10}\cline{12-14}
\multicolumn{1}{c}{}                                                                   &                           & \multicolumn{1}{c}{$\alpha$} & 0.01                 & 0.05                 & 0.05                 &                      & 0.01                 & 0.05                 & 0.10                            &                      & 0.01                 & 0.05                 & 0.10                  \\ 
\hline
\multicolumn{1}{c}{3}                                                                  & (10, 10)                  &                              & 0.7178               & 0.8945               & 0.9425               & \multicolumn{1}{c}{} & 0.5691               & 0.7211               & 0.8019                          &                      & 0.4738               & 0.6753               & 0.7629                \\
                                                                                       & (20, 20)                  &                              & 0.9629               & 0.9928               & 0.9974               &                      & 0.8645                     &    0.9680                  &  0.9880                               &                      &  0.8340                    &  0.9605                    &   0.9845                    \\
                                                                                       & (25, 20)                  &                              & 0.9773               & 0.9971               & 0.9992               &                      &  0.9125                    &  0.9760                    &  0.9925                               &                      &  0.8945                     & 0.9705                      &  0.9890                     \\
                                                                                       & \multicolumn{1}{l}{}      &                              & \multicolumn{1}{l}{} & \multicolumn{1}{l}{} & \multicolumn{1}{l}{} &                      & \multicolumn{1}{l}{} & \multicolumn{1}{l}{} & \multicolumn{1}{l}{}            &                      & \multicolumn{1}{l}{} & \multicolumn{1}{l}{} & \multicolumn{1}{l}{}  \\
\multicolumn{1}{c}{5}                                                                  & (10, 10)                  &                              & 0.8965               & 0.9738               & 0.9868               &                      & 0.7977               & 0.9042               & 0.9420                          &                      & 0.6976               & 0.8724               & 0.9237                \\
                                                                                       & (20, 20)                  &                              & 0.9980               & 0.9999               & 1.0000               &                      & 0.9910                     & 0.9985                     &   0.9990                              &                      &  0.9875                    &       0.9985               &   0.9990                    \\
                                                                                       & (25, 20)                  &                              & 0.9991               & 1.0000               & 1.0000               &                      &   0.9955                   &  0.9995                    &   1.0000                               &                      &   0.9935                   &   0.9995                   & 1.0000                      \\
                                                                                       & \multicolumn{1}{l}{}      &                              & \multicolumn{1}{l}{} & \multicolumn{1}{l}{} & \multicolumn{1}{l}{} &                      & \multicolumn{1}{l}{} & \multicolumn{1}{l}{} & \multicolumn{1}{l}{}            &                      & \multicolumn{1}{l}{} & \multicolumn{1}{l}{} & \multicolumn{1}{l}{}  \\
\multicolumn{1}{c}{7}                                                                  & (10, 10)                  &                              & 0.9476               & 0.9881               & 0.9934               &                      & 0.8759               & 0.9432               & 0.9647                          &                      & 0.7701               & 0.9212               & 0.9545                \\
                                                                                       & (20, 20)                  &                              & 0.9999               & 1.0000               & 1.0000               &                      & 0.9985                     & 1.0000                     &    1.0000                             &                      & 0.9980                      & 1.0000                      & 1.0000                       \\
                                                                                       & (25, 20)                  &                              & 1.0000               & 1.0000               & 1.0000               &                      & 0.9992                      &  1.0000                    &  1.0000                               &                      &  0.9990                     & 1.0000                      & 1.0000                         \\
\hline
\end{tabular}
\end{table}

{\bf Scenario II:} Let $X$ follow standard exponential distribution and $Y$ follow the Gumbel distribution with scale parameter $\gamma$ with the cumulative distribution function $F(x)=\exp{\left(-e^{-\gamma x}\right)}$. Note that $\frac{r_Y(x)}{r_{X}(x)} \text{ is an increasing function of } x>0$ if and only if $\gamma> 1$. The considered values of $\gamma$ are 3, 5 and 7. We generate two sets of independent random observations of size $m$ and $n$ respectively from the exponential distribution with mean $1$ and the Gumbel distribution with scale parameter $\gamma$. The empirical power of the tests based on $U_{MW}$, $U_{JEL}$ and $U_{AJEL}$ are calculated based on 10,000 replications with sample sizes $(m,n)=(10, 10), (20, 20) \text{ and } (25, 20)$. We present the computed empirical power in Table \ref{powgum} for 1\%, 5\% and 10\% levels of significance. From Table \ref{powgum}, we observe that all the proposed tests have good power for the parameters when $m\geq 20$ and $n\geq 20$. The $U_{MW}$ test performs marginally better than $U_{JEL}$ for small values of $m$ and $n$. Moreover, $U_{JEL}$ has also slightly good power for the parameters we consider while comparing with the test $U_{AJEL}$ for small values of $m$ and $n$. Finally, as expected the empirical power increases when the value of $\gamma$ increases.

\section{Analysis of real data sets}
\label{sec4}
Before going to analyse real data sets, it is important to present graphical analysis for the PRHR assumption. Let $F_{0m}$ and $F_n$ represent empirical distribution functions based on the samples $X_1, X_2, \dots ,X_m$ of size $m$ and $Y_1, Y_2, \dots ,Y_n$ of size $n$ from $F_0$ and $F$ respectively. If the plots of $\log{\left(-\log{\left(F_n(t)\right)}\right)}$ and $\log{\left(-\log{\left(F_{0m}(t)\right)}\right)}$ exhibit almost same difference then the data may indicate proportionality in reversed hazard rate. In the present paper, this plot will be referred to as log-log plot. In the next subsections, two real data sets have been analysed in the sequel.

\subsection{Data in the context of  brain injury-related biomarkers}
\begin{figure}[ht]
    \centering
    \includegraphics[scale=.72]{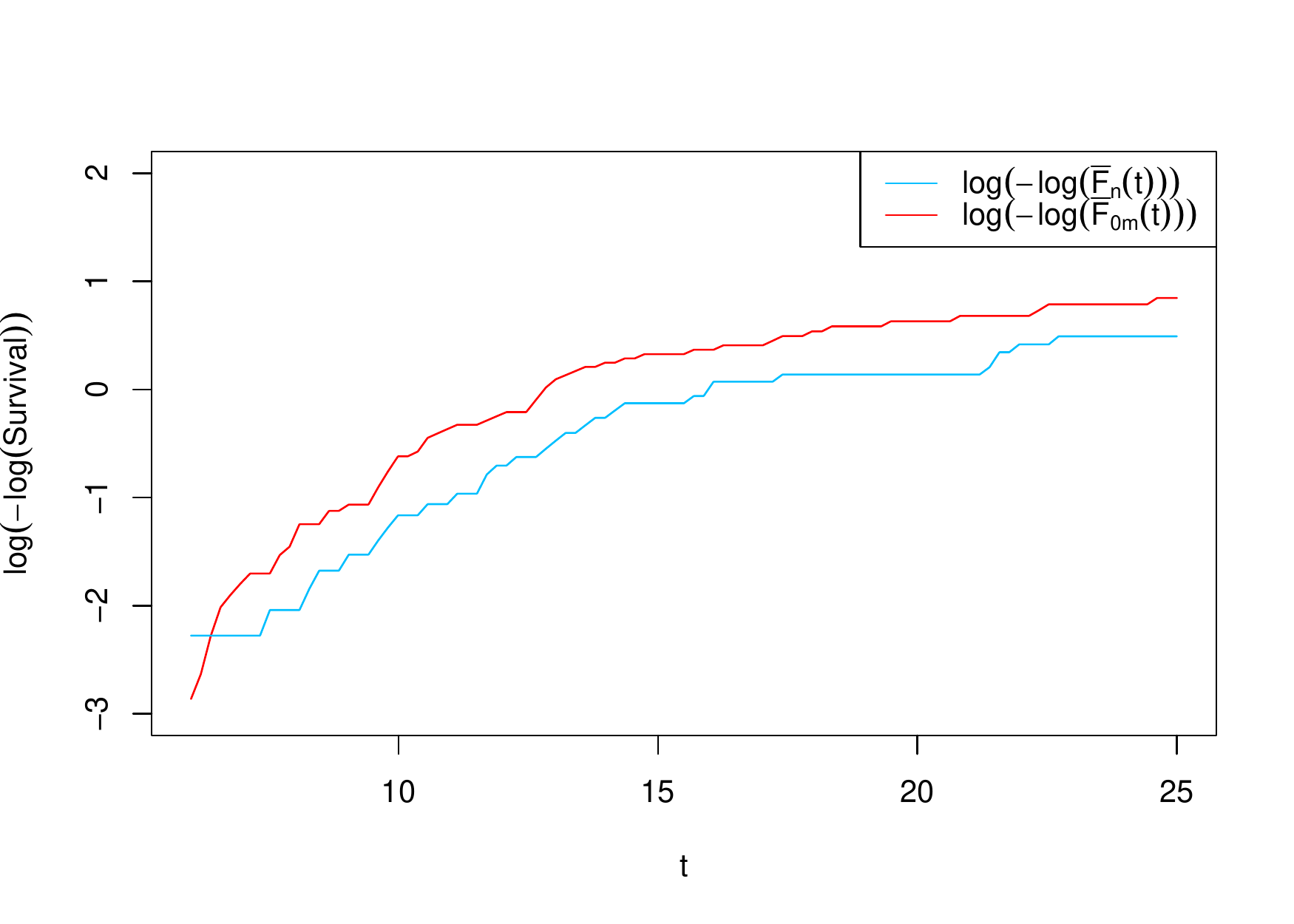}
    \caption{log-log plot of the empirical distribution functions based on the aSAH data} 
    \label{asah}
\end{figure}

 \cite{turck2010multiparameter} conducted a study for outcome prediction following aneurysmal subarachnoid hemorrhage (aSAH) using a combination of clinical scores together with brain injury-related biomarkers of 113 patients admitted within 48 hours. After six months, based on the the condition of patients, the outcome was categorised as good when the Glasgow Outcome Scale
(GOS) was greater than equal to $4$ (41 observations) or poor (72 observations) otherwise. Here, the score of a most important marker, i.e, nucleoside diphosphate kinase A (NDKA) level is considered. Figure \ref{asah} indicates in favour of the proportionality in the reversed hazard rate between two groups. The values of test statistics $U_{MW}$, $U_{JEL}$ and $U_{AJEL}$ have been calculated and the corresponding $p$-values are reported in Table \ref{musculardata}. Here all the tests fail to reject the null hypotheses $H_0$ at 5\% level of significance which is in accordance with the graphical representation given in Figure \ref{asah}. 

\begin{table}
\centering
\caption{Summary statistics for the aSAH data}
\begin{tabular}{cclcclcc} 
\hline
\multicolumn{2}{c}{$U_{MW}$}                                                                      &  & \multicolumn{2}{c}{$U_{JEL}$}                                                            &  & \multicolumn{2}{c}{$U_{AJEL}$}                                                            \\ 
\cline{1-2}\cline{4-5}\cline{7-8}
\begin{tabular}[c]{@{}c@{}}Test statistic\\~value\end{tabular} & $p$-value                       &  & \begin{tabular}[c]{@{}c@{}}Test statistic \\value\end{tabular} & $p$-value              &  & \begin{tabular}[c]{@{}c@{}}Test statistic \\value\end{tabular} & $p$-value               \\ 
\hline
\multicolumn{1}{l}{1.214275}                                   & \multicolumn{1}{l}{0.1123214} &  & \multicolumn{1}{l}{1.43033}                                           & \multicolumn{1}{l}{0.2317105} &  & \multicolumn{1}{l}{1.371299}                                           & \multicolumn{1}{l}{0.2415888}  \\
\hline
\end{tabular}
\end{table}

\subsection{Ducheme muscular dystrophy data}
Ducheme muscular dystrophy (DMD) is a well-known
muscular dystrophy and is genetically transmitted from a mother to her child.  It is known as rapid progression of muscle degeneration in early life and there is no cure at all for this disease. Thus, the diagnosis of affected females is very important. \cite{andrews2012data} reported some data in Table 38.1 which was collected during a program conducted at a hospital for sick children in Toronto. In our study, we consider one of the serum enzyme levels, i.e., creatine kinase (CK), for 75 carriers and 134 noncarriers (healthy females). The following Figure \ref{ckprhr} indicates in favour of the increasing ratio of reversed hazard rates between two groups.  The values of test statistics $U_{MW}$, $U_{JEL}$ and $U_{AJEL}$ have been calculated and the corresponding $p$-values are presented in Table \ref{musculardata}. The table suggests to reject the null hypothesis $H_0$ at 5\% level of significance. In this context, one should also note that these findings are in agreement with the graphical analysis. 

\begin{figure}[ht]
    \centering
    \includegraphics[scale=.72]{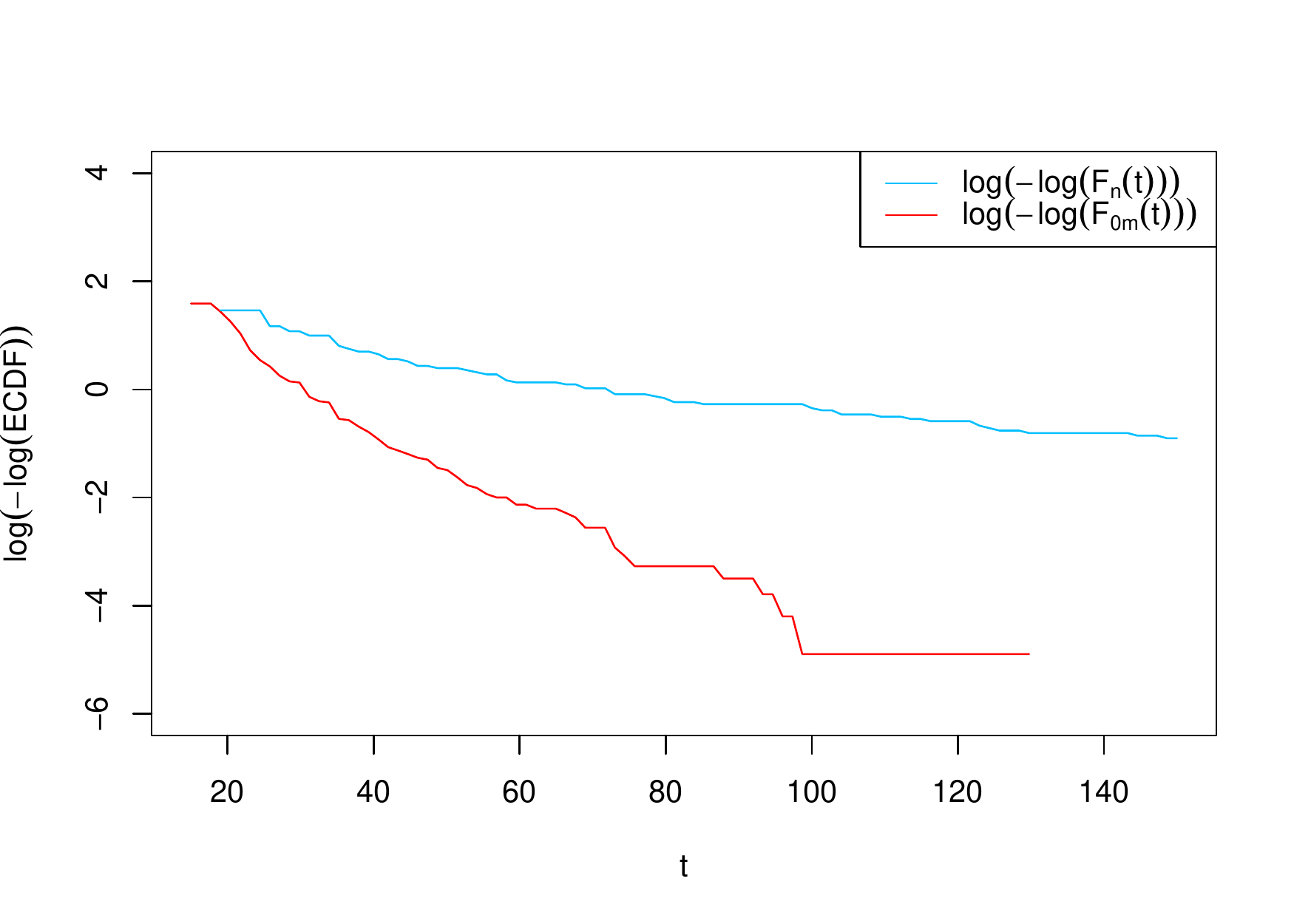}
    \caption{log-log plot of the empirical distribution functions based on the DMD data}
    \label{ckprhr}
\end{figure}

\begin{table}[ht]
\centering
\caption{Summary statistics for the DMD data}
\label{musculardata}
\begin{tabular}{cclcclcc} 
\hline
\multicolumn{2}{c}{\begin{tabular}[c]{@{}c@{}}\\$U_{MW}$\end{tabular}}        & \multicolumn{1}{c}{} & \multicolumn{2}{c}{$U_{JEL}$}                                              & \multicolumn{1}{c}{} & \multicolumn{2}{c}{$U_{AJEL}$}                                              \\ 
\cline{1-2}\cline{4-5}\cline{7-8}
\begin{tabular}[c]{@{}c@{}}Test statistic\\~value\end{tabular} & $p$-value    &                      & \begin{tabular}[c]{@{}c@{}}Test statistic \\value\end{tabular} & $p$-value &                      & \begin{tabular}[c]{@{}c@{}}Test statistic \\value\end{tabular} & $p$-value  \\ 
\hline
4.96845                                                        & 3.374503e-07 &                      &  62.93436                                                              &   2.109424e-15        &                      &  59.5923                                                              &       1.165734e-14     \\
\hline
\end{tabular}
\end{table}

\section{Discussion}
\label{sec5}
In this paper, a two-sample nonparametric test based on two independent samples is proposed for verifying the PRHR assumption. Based on a consistent U-statistic three statistical methodologies have been developed exploiting U-statistics theory, JEL and AJEL method.  A simulation study has been performed to assess the merit of the proposed test procedures. %Empirical Type I error and  empirical power for Scenario I have been computed based on 1,000 replications to reduce the computation time. However, empirical power for Scenario II has been calculated based on 10,000 replications where the computation time for each row of Table \ref{powgum} takes about 4.5 days. Finally, the test is applied to a data in the context of brain injury-related biomarkers and a data related to Ducheme muscular dystrophy. 

One can observe that the testing problem, given in (\ref{testprob}),  does not have the flexibility of handling the nonmonotonic behaviour of the ratio of two hazard rates. There  is  scope  for further work in this direction. Another possible avenue of future research is that the testing problem considered in this paper can be extended to the case of left censored data since \cite{kalbfleisch1989inference} shows the importance of RHR  in  the estimation of the survival function under left censored data.

\section*{Acknowledgements}
 The present author is grateful to the Theoretical Statistics and Mathematics Unit of Indian Statistical Institute, Delhi Centre, India for a visiting position and providing necessary infrastructure.

\bibliographystyle{apa}
\bibliography{winnower_template}
\appendix

\end{document}